\documentclass[sigconf,authorversion]{acmart}

\usepackage{booktabs} 

\usepackage{proof}
\usepackage{graphicx}

\setcopyright{rightsretained}

\acmConference[]{}{}{}

\begin{document}
\title{Languages of Play}
\subtitle{Towards semantic foundations for game interfaces}

 \author{Chris Martens}
 \affiliation{%
   \institution{North Carolina State University}
   \city{Raleigh} 
   \state{NC} 
 }
 \email{martens@csc.ncsu.edu}
 
 \author{Matthew A. Hammer}
 \affiliation{%
   \institution{University of Colorado Boulder}
   \city{Boulder}
   \state{CO}
 }
 \email{matthew.hammer@colorado.edu}

\begin{abstract}
Formal models of games help us account for and predict behavior, leading to
more robust and innovative designs. While the games research community has 
proposed many formalisms for both the ``game half'' (game models, game
description languages) and the ``human half'' (player modeling) of a game
experience, little attention has been paid to the {\em interface} between
the two, particularly where it concerns the player expressing her intent
toward the game. We describe an analytical and computational toolbox based
on programming language theory to examine the phenomenon sitting between
control schemes and game rules, which we identify as a distinct {\em
player intent language} for each game.
\end{abstract}


\maketitle

\section{Introduction}

To study how players interact with games, we examine both the rules of the
underlying system and the choices made by the player. 
The field of player modeling has identified the
value in constructing models of player cognition: while a game as
a self-contained entity can allow us to learn about its mechanics and
properties as a formal system, we cannot understand the {\em dynamics} of
that system unless we also account for the human half of the equation.
Meanwhile, Crawford~\cite{crawford2003chris} identifies the necessity of
looking at the complete information loop created between a player and a
digital game, defining {\em interactivity} in games as their ability to
carry out a conversation with a player, including listening, processing,
and responding, identifying the importance of all three to the overall
experience.

\begin{figure*}
\includegraphics[width=0.75\textwidth]{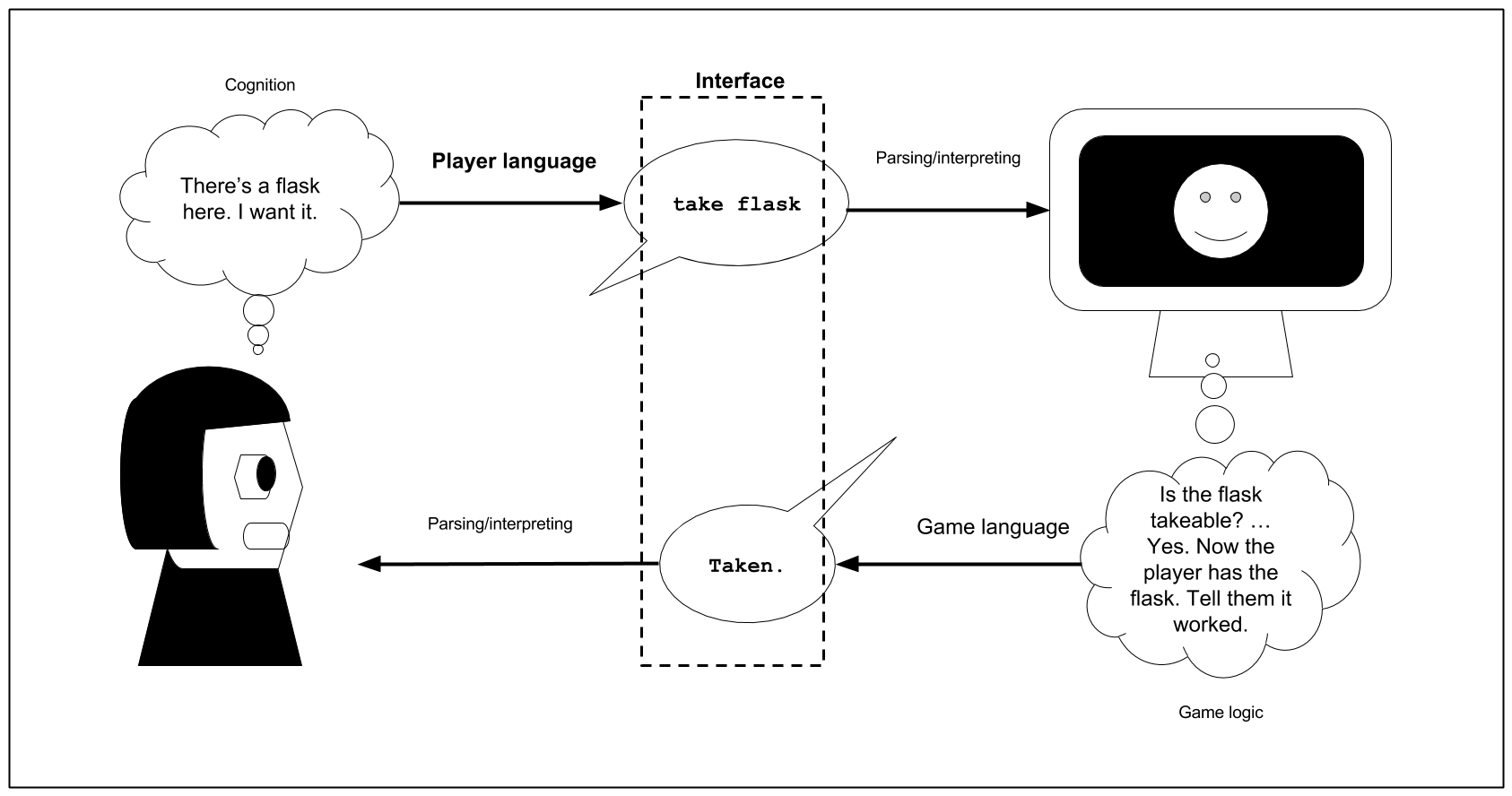}
\caption{A process diagram of a game loop: player and game conversation as
it relates to language, interface, and cognition.}
\label{fig:gameloop}
\end{figure*}

Given this understanding of games-as-conversation, we should expect to
discover something like a {\em language} through which games and players converse. 
In Figure~\ref{fig:gameloop}, we illustrate the game-player loop as a
process which includes an {\em interface} constituting such a language.
The Game Ontology Project~\cite{zagal2007towards} describes game interfaces
as follows:
\begin{quote} \em
The interface is where the player and game meet, the mapping
between the embodied reactions of the player and the manipulation of game
entities. It refers to both how the player interacts with the game and how
the game communicates to the player.
\end{quote}
The first part, {\em how the player interacts with the game}, is
called the {\em input}, which is further subdivided into {\em input device}
and {\em input method}. Input devices are hardware controllers (mice,
keyboards, joysticks, etc.) and input {\em methods} start to brush the
surface of something more semantic: they include choices about {\em locus of
manipulation} (which game entities can the player control?) and direct
versus indirect action, such as selecting an action from a menu of options
(indirect) versus pressing an arrow key to move an avatar (direct).

However, any close look at interactive fiction, recent mobile games, or
rhythm games (just to name a few examples) will reveal that design choices
for input methods have much more variety and possibility than these two
dimensions. In this paper, we propose a framework to support analyzing and exploring
that design space. 

Our first step is to refine input {\em methods} to input {\em languages}:
we are ultimately asking, how can a player communicate their intent, and
how does a digital game recognize this intent?  So, in linguistic terms,
the ``phonemes'' of such a language are hardware controls such as button
presses and joystick movement. Then, the syntax and semantics are defined
by each game individually, depending on what meaning they give to each
control input. This language defines the verbs of the game, which may
include moving, selecting inventory items, examining world items, applying
or using items, entering rooms, and combat actions.  (Note that such a
language is also distinct from a game's {\em mechanics}: mechanics include
system behavior which is out of the player's control, such as falling with
gravity, non-player character actions, and other autonomous behavior.) This
language is both {\em afforded} by the game designer---she must communicate
to players which verbs are available---and {\em constrained} by her---she
may declare certain expressions invalid.

Since the constraints on such a language are wholly determined
by a piece of software (the game interface), we argue that it
has more in common with a {\em programming language}\footnote{We define
  programming languages broadly as formal languages whose meaning is fully
grounded in a computational system.} (PL) than a natural language.
Accordingly, each game in some sense {\em defines its own programming
language}. In a slogan, we could term this project {\em games as
programming languages}.
Specifically, we propose \emph{player intent languages} as a PL-inspired
framework for designing player-game interfaces.


This analogy opens up a whole field of methodology to try applying to
games.  The PL research community has a long and deep
history of assigning mathematically formal semantics to languages and
analyzing those semantics. As games researchers become more interested in
the emergent consequences of the systems they assemble, the tools of PL
theory have a lot to offer. For example, PL theory provides an account of
{\em compositionality}, i.e. how fragments of expression fit together to
form higher-level meanings. In games, this translates into being able to
understand player {\em skills} or strategies as compositions of player
actions, which we demonstrate in this paper by using a formalized input
language as a kind of ``player AI scripting language.''



%
%

Furthermore, by considering a player's language of expression as an object
of study in its own right, we center them as a co-designer of the experience
afforded by a game. When we treat a player's interactions as not simply an
arbitrary sequence of button presses that advances and reveals the
designer's intent, but instead as its own distinct {\em voice} that a game
system must listen and respond to, we enable the player to {\em co-create}
with the system, potentially developing deeper systems understanding
and emotional investment.

In this paper, we propose \emph{player intent languages}, a
programming languages-based approach to designing player-game
interfaces as formal objects.
In the remainder of the paper, we tour this approach through concrete
examples.
Specifically, we consider a simple game design space and make points
in this space precise by introducing the components of a programming
language:
abstract syntax (Section~\ref{sec:syntax}),
type system~(Section~\ref{sec:typesys}), 
and operational semantics~(Section~\ref{sec:opsem}).
For each, we give a corresponding concept in the game world.
By grounding these game concepts in analogous programming language
concepts, we gain powerful PL reasoning tools and design methodologies
to benefit the game design process.

We demonstrate the payoff of this line of thought by extending the
metaphor with \emph{play traces} as \emph{straight-line
  programs} (Section~\ref{sec:traces}), and {\em player skills} as
\emph{general programs} (e.g., programs with parameters, branching and
looping)~(Section~\ref{sec:skills}).
These structures give semantic logs and general strategies,
respectively, for accomplishing a task in the game world.
The framework of player intention languages gives rise to further research
directions, which we briefly outline and discuss before concluding
(Sections~\ref{sec:discussion} and \ref{sec:conclusion}).


\section{Related Work}

Cardona-Rivera and Young~\cite{cardona2014games}
detailed a conceptual framework following the slogan {\em games as
conversation}, grounding the communicative strategies of games in cognitive
science for human-to-human conversational understanding, such as Grice's
maxims~\cite{grice1975logic}. They offer a linguistic and semiotic approach to
understanding how a game communicates affordances (possibilities for
action) to a player.  For an account of the game's half of the equation,
which includes the visual, textual, and audio feedback mechanisms intended
to be processed by the player, this application of linguistics, psychology,
and design seems appropriate, much like the study of cinematic language for
film.  On the other hand, we argue that a PL approach better supports
understanding of the player-to-game direction, since the language the
player speaks toward a digital game is formal and unambiguous.

Researchers have previously recognized the value in formalizing interaction
vocabularies, realizing certain interaction conventions as a {\em single}
``video game description language''~\cite{ebner2013towards} whose
implementation as VGDL~\cite{schaul2013video} has been used in game AI research. We
suggest instead that the design space of player languages is as varied as
the design space of programming languages and herein give an account of
what it would mean to treat each language individually.
Our project suggests that an appropriately expressive computational
framework analogous to VGDL should be one that can accommodate the encoding
of many such languages, such as a {\em meta-logical framework} like the
Twelf system for encoding and analyzing programming language
designs~\cite{pfenning1999system}. 

Any investigation into formalizing actions within an interactive system
shares ideas with ``action languages'' in AI extending as far back as
McCarthy's situation calculus~\cite{mccarthy1969some} and including
planning languages and process calculi.  These systems have been studied in the
context of game design, e.g. the Ludocore system~\cite{smith2010ludocore};
however, AI researchers are mainly interested in these formalisms as internal
representations for intelligent systems and the extent to which they support
reasoning.  Conversely, we are interested their potential to support player
expression and facilitate human-computer conversation.

Some theoretical and experimental investigations have been carried out
about differences between game interfaces along specific axes, such as
whether the interface is ``integrated'' (or one might say diagetic), versus
extrinsic to the game world in the form of menus and
buttons~\cite{llanos2011players, jorgensen2013gameworld}. These
investigations suggest an interest in more detailed and formal ontologies
of game interfaces, which our work aims to provide.

From the PL research side, we note existing efforts to apply PL
methodology to user interfaces, specifically in the case of program
editors.
Hazelnut is a formal model of a program editor that enforces that
every edit state is meaningful (it consists of a well-defined syntax
tree, with a well-defined type)~\cite{omar17hazelnut}.
Its type system and editing semantics permit \emph{partial programs},
which contain missing pieces and well-marked type inconsistencies.
Specifically, Hazelnut proposes a \emph{editing language}, which
defines how a cursor moves and edits the syntax tree; the planned
benefits of this model range from better editing assistance, the
potential to better automate systematic edits, and further
context-aware assistance and automation based on statistical analysis
of (semantically-rich) corpora of recorded past edits, which consist
of \emph{traces} from this language~\cite{omar17hazel}.
Likewise, in the context of game design, 
we expect similar benefits from the lens of language design.

%

\section{A Framework for \\ Player Intent Languages}

\newcommand{\param}[1]{\langle #1 \rangle}
\newcommand{\syn}[1]{\mathsf{#1}}

  In the formal study of a programming langugage, one may define a language
  in three parts: syntax, type system, and operational semantics.
  \begin{itemize}
    \item The {\em syntax} is written in the form of a (usually)
      context-free grammar describing the allowable expressions. One
      sometimes distinguishes between {\em concrete syntax}, the literal
      program tokens that the programmer strings together in the act of
      programming, and {\em abstract syntax}, the normalized ``syntax
      tree'' structures that ultimately get interpreted. 
    \item An {\em operational semantics} defines how runnable programs
      (e.g. a function applied to an argument) {\em reduce} to values. This
      part of the definition describes how actual computation takes place
      when programs in the language are run. It is important to note that
      the operational semantics need not reflect the actual {\em
      implementation} of the language, nor is it specific to a ``compiled''
      versus ``interpreted'' understanding of the language: it is simply a
      mathematical specification for how any compiler or interpreter for
      the language should behave.
    \item A {\em type system} further refines the set of syntactically
      valid expressions into a set of {\em meaningful} expressions, and provides
      a mapping between an expression and an approximation of its meaning.
      Type systems are usually designed in conjunction with the operational
      semantics to have the property that every expression assigned a
      meaning by the type system should have a well-defined runtime
      behavior. In practice, however, type systems can only approximate
      this correspondence. Some err on the more permissive side--e.g.
      C's type system will permit invalid memory accesses with no
      language-defined behavior--and some err on the more restrictive side,
      e.g. Haskell's type system does not permit any untracked
      side-effects, at the expense of easily authoring e.g. file input/output
      (without first learning the details of the type system).
  \end{itemize}
  
  Providing a formal language definition in programming languages research
  has several purposes. One is that it enables researchers to explore
  and prove formal properties of their language, such as {\em well-typed
  programs don't go wrong}, or in a language for concurrency, a property
  like deadlock freedom. However, an even more crucial advantage of a
  language specification is not mathematical rigor but human capacity to do
  science. A language definition is a {\em specification}, similar to an
  application programmer interface (API) or an IEEE standard: it describes
  an unambiguous interface to the language along an {\em abstraction
  boundary} that other human beings may access, understand, and implement,
  without knowing the internals of a language implementation.  It is a
  necessary component of reproducibility of research, and it allows
  researchers to build on each other's work. We believe that an embrace of
  formal specification in games research can play a similarly important
  function.

  \begin{table}
  \begin{tabular}{l|ll}
    PL concept & Game concept\\
    \hline
    Syntax & Recognized player intents & (Section \ref{sec:syntax})\\
    Operational semantics & Game mechanics & (Section \ref{sec:opsem})\\
    Type system & Contextual interface & (Section \ref{sec:typesys})\\
    Straight-line programs & Play traces & (Section \ref{sec:traces})\\ 
    General programs & Player skills & (Section \ref{sec:skills})
    \\
    \hline
  \end{tabular}
  \caption{Player intent languages: 
    \\
    Formal decomposition (left) and correspondances (right).}
  \label{tab:correspondence}
  \end{table}

  Having provided loose definitions of these terms, we now wish to draw out
  the analogy between a {\em language} specification and a {\em game}
  specification. To treat a game in this manner, we wish to consider player
  affordances and actions, as well as their behavior (mechanics) in the
  context of the game's running environment. We summarize the components of
  this correspondence in Table~\ref{tab:correspondence}.

  \newcommand{\cmove}{\mathsf{move}}
  \newcommand{\ctake}{\mathsf{take}}
  \newcommand{\ccollect}{\mathsf{collect}}

\begin{figure*}
    \begin{tabular}{cc}
      \includegraphics[height=0.2\textheight]{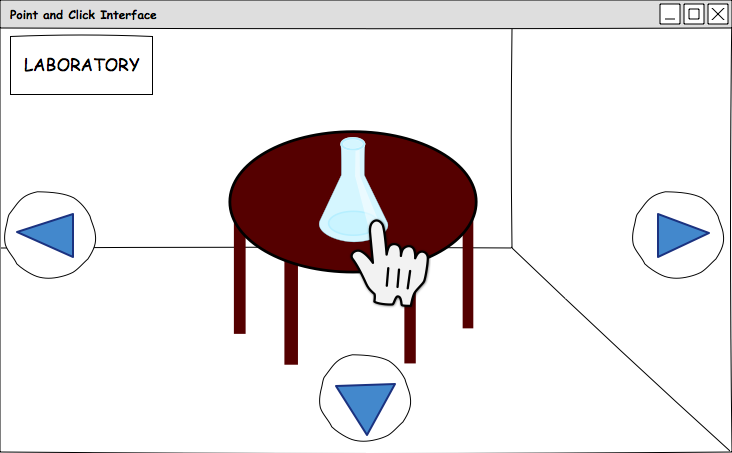} &
    \includegraphics[height=0.2\textheight]{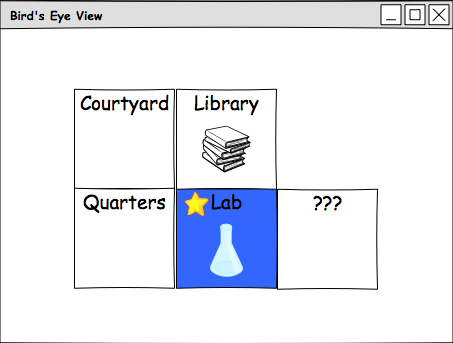}
    \\
    Point-and-click & Bird's eye view/WASD+\\
    \includegraphics[height=0.2\textheight]{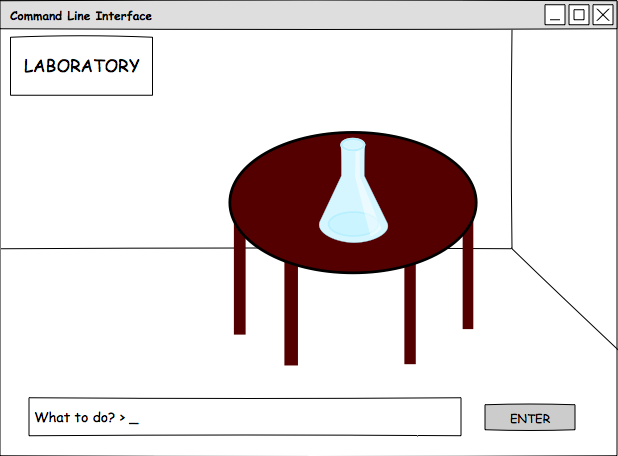} &
    \includegraphics[height=0.2\textheight]{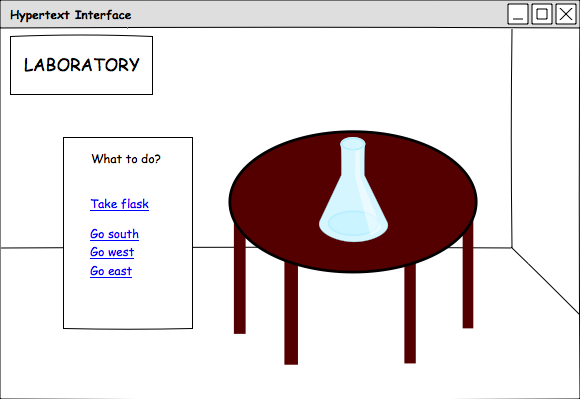}\\
    Command-line & Hypertext
  \end{tabular}
  \caption{Four different user interfaces for the move/take game.}
  \label{fig:uis}
  \end{figure*}

  We will use as a running example a minimal virtual environment with two
  player actions: (1) movement through a discrete set of rooms in a
  pre-defined map ($\cmove$); (2) acquiring objects placed in those rooms to store in
  a player inventory ($\ctake$).  We consider five (somewhat aribitrary)
  possibilities in the design space of interfaces for such a game,
  summarized visually in Figure~\ref{fig:uis}:

  \begin{itemize}
   \item {\bf Point-and-Click:} A first-person viewpoint interface where the
     meaning of each click is defined based on the region the cursor falls
     in. Clicking near any of the four screen edges moves in that
     direction; clicking on a sprite representing an item takes it.
  \item {\bf Bird's-Eye:} A top-down viewpoint interface where the player can see multiple
    rooms at once, and can click on rooms and objects that are far away,
    but those clicks only do something to objects in the same room or
    adjacent rooms.
  \item {\bf WASD+:} A keyboard or controller-based interface with
    directional buttons (e.g. arrow keys or WASD) move an avatar in the
    correspondingi direction, and a separate key or button expresses
    the $\ctake$ action, which takes any object in the same room. (This
    interface may be used for either of the two views described above.)
  \item{\bf Command-Line:} The player interacts by typing free-form text, 
    which is then parsed into commands, such as \verb|take lamp| and \verb|move north|.
  \item{\bf Hypertext}: A choice-based interface where all available options are
    enumerated as textual links from which the player chooses.
  \end{itemize}

  In the following sections, we will consider these possibilities in light
  of design choices relevant to the specified aspect of PL design.

  \section{Player intent as syntax}
  \label{sec:syntax}

  The {\em syntax} of a game is its space of recognized player intentions.
  Note that {\em intention} is different from {\em action} in the sense
  that we don't necessarily expect each well-formed intention to change
  anything about the game state: a player can intend to move north, but if
  there is no room to the north of the player when she expresses this
  intent, no change to the game's internal state will occur. Nonetheless,
  depending on the design goals of the game, we may wish to recognize this
  as a valid intent so that the game may respond in some useful way (e.g.
  with feedback that the player cannot move in that direction).
  
  In our example game, the choice of syntax answers questions such as: can
  the player click anywhere, or only in regions that have meaning? Can the
  player type arbitrary commands, or should we provide a menu or
  auto-complete text so as to prevent the player from entering meaningless
  commands?  In PL, we can formalize these decisions by describing an {\em
  abstract syntax} for our language, which is typically assumed to be
  context-free and thus specified as a Bachus-Naur Form (BNF) grammar. Our
  examples below follow the interfaces shown visually in
  Figure~\ref{fig:uis}.
  
  \textbf{WASD+ Interface}:
  One way of writing the BNF for the WASD+ interface is:
  \begin{eqnarray*}
  direction &::=& \mathsf{north} \mid \mathsf{south} \mid \mathsf{east}
    \mid \mathsf{west} \\
  intent &::=& \cmove \langle direction \rangle \mid \ccollect
  \end{eqnarray*}
  The hardware interface maps onto this syntax quite directly:
  each arrow key maps onto a $\cmove$ action in the corresponding
  direction, and the specified other key maps onto $\ccollect$.

  \textbf{Bird's Eye View Mouse Interface}:
  On the other hand, a clicking-based interface to a top-down map could
  enable the player to click on any room on the map and any item within a
  room. This syntax would look like:
  \begin{eqnarray*}
    room   &::=& \syn{courtyard} \mid \syn{library} \mid \syn{quarters}
              \mid \syn{lab}\\
    item   &::=& \syn{flask} \mid \syn{book}\\
    entity &::=& \syn{room} \mid \syn{item}\\
    intent &::=& \syn{click}\param{entity}
  \end{eqnarray*}
  Note that this syntax, compared to that for WASD+, describes a larger set
  of possible utterances, even though it has the exact same set of
  permitted game behaviors (a player may only move into adjacent rooms and
  take items that share a room with them). 


  \textbf{Command Line Interface}:
  The command-line interface would have an even larger space of expressible
  utterances if we consider all typed strings of characters to be valid
  expressions, but that syntax is too low-level for linguistic
  considerations. Supposing we interpose a parsing layer between arbitrary
  typed strings and syntactically-well-formed commands, we can define the
  abstract syntax as follows (where $direction$ and $item$ are assumed to
  be defined as they were in the previous examples):
  \begin{eqnarray*}
    intent &::=& \cmove\param{direction} \mid \ctake\param{item}
  \end{eqnarray*}
  Assuming the player ``knows the language,'' i.e. knows that $\cmove$
  and $\ctake$ are valid commands, and in fact the {\em only} valid
  commands, and assuming that she knows how to map the visual affordances
  (e.g. image of the flask) to the typed noun (e.g. \verb|flask|), the
  experience afforded by this interface is quite similar to the WASD+
  interface. The main difference is that the player must specify an {\em
  argument} to the $\syn{take}$ command, asking the player to formulate a
  more complete (and unambiguous) intent by actually naming the object she
  wishes to take.


  \textbf{Hypertext interface}:
  Finally, we consider the intent language for the hypertext interface.
  This is one of the most difficult interfaces to formulate in linguistic
  terms, because it either requires that we formalize link selection in an
  acontextual way (e.g. as a numeric index into a list of options of
  unknown size) or that we formulate each link {\em from each page} as its
  own separate command, each of which has meaning in only one specific game
  context (namely, when the player is on the page containing that link).
  The former feels like a more general formulation of hypertext that is not
  relevant to any particular game, and since we are aiming to provide a
  correspondence between specific games and languages, we opt for the
  latter:
  \begin{eqnarray*}
    intent &::=& \syn{select}\param{choice}\\
    choice &::=& \syn{take\_flask\_from\_lab}\\
           &\mid& \syn{take\_book\_from\_library}\\
           &\mid& \syn{go\_south\_from\_lab} \\
           &\mid& \syn{go\_east\_from\_lab} \\
           &\mid& \syn{go\_west\_from\_lab} \\
           &\mid& \syn{go\_north\_from\_library} \\
           &\mid& \hdots
  \end{eqnarray*}
  Some hypertext authors put a lot of effort into scaffolding the
  choice-based experience with a richer language, e.g. by repeating the
  same set of commands that behave in consistent ways across different
  pages, or by creating menu-like interfaces where text cycles between
  options on an otherwise static page. In this way, hypertext as a medium
  might be said as providing a platform for designers to create their own
  interface conventions, rather than relying on a set of pre-established
  ones; by the same token, hypertext games created by inexperienced
  interface (or language) designers may feel to players like being asked to
  speak a foreign language for each new game.

  \subsection*{Additive and subtractive properties of syntax}
  
  By now we are able to observe that, just like the rest of a game's rules,
  its syntax has both additive and subtractive properties. It provides the
  menu of options for which hardware interactions are {\em relevant}, i.e.
  likely to result in meaningful interaction with the game system, but it
  also establishes which utterances within that set are {\em disallowed},
  or ill-formed---e.g. that it is not meaningful to say ``take'' without
  providing an object to the command, or that ``take north'' is ill-formed.

  Correspondingly, an important decision that impacts game design is (a)
  how discoverable the additive affordances are (e.g. can the player
  determine that ``examine'' is a meaningful verb without already
  possessing literacy in the game's genre?) and (b) the extent to which the
  user interface makes meaningless expressions impossible to form. For
  example, in a hypertext interface, all links lead somewhere---so every
  intent the player can form, i.e. clicking a link on the page, will get a
  valid response from the game, whereas ``take fnord'' typed at a
  command-line interface may be recognized by the parser, but meaningless
  to a game where ``fnord'' is not a noun.  Decisions about these two
  (related) dimensions will determine the extent to which {\em learning the
  language}, an exploratory but sometimes frustrating process, is a central
  challenge of the game.

  \section{Game mechanics as \\ operational semantics}
  \label{sec:opsem}



%

  \newcommand{\OmitThis}[1]{}

Consider the right-hand-side of Figure~\ref{fig:gameloop}: The game
parses and interprets an unambiguous syntax of player intent, which
either advances the game state (as shown in the figure), or the game
state cannot advance as the player intended, which the game somehow
signals to the player (not shown).

We model a \emph{response} to the move-take player intent with the
following BNF definition, of~$\mathit{resp}$:
  \begin{eqnarray*}
  \mathit{resp} &::=& \mathsf{success} \mid \mathsf{failure}
  \end{eqnarray*}
Figure~\ref{fig:gameloop} shows the case where the player intent of
``take flask'' (formally, the syntax ``$\ctake$'') leads to the game
world performing this intent as a successful action, and responding
accordingly with ``Taken.''
Formally, we model this $\mathit{resp}$ as~``$\mathsf{success}$,'' as
defined above.
Likewise, if the flask cannot be taken (e.g., the flask is not near
the player, or is already in the player's possession, etc.), the game
responds with~$\mathsf{failure}$.

As with the player's intent, which may exist as both raw input
\emph{and} as formal syntax, each formal response can be conveyed as
raw output in a variety of ways (e.g., as textual words, pictures, or
sounds).
In real games there are often two two levels of game-to-player
feedback:
Feedback through the game world, and feedback \emph{outside} the game
world (e.g., a pop-up message with an error, guidance or advice).
For simplicity, move-take gives feedback outside of the game state,
e.g., as pop-up messages.


\newcommand{\GameStep}[4]{
 \left< #1; #2 \right>
 \longrightarrow 
 \left< #3; #4 \right>
}

To capture the formal relationship between player intent as syntax,
and game response as syntax, we introduce a four-place
\emph{game-step} relation:
\[
\GameStep{G_1}{\mathit{intent}}{G_2}{\mathit{resp}}
\]
This relation formalizes the dynamic behavior of the right-hand-side
of Figure~\ref{fig:gameloop}.  It consists of four parts: An initial
game state~$G_1$, a player intent~$\mathit{intent}$, a resulting game
state~$G_2$ and a game response~$\mathit{resp}$.

As is standard in PL formalisms, we give the rules that define this
relation as inductive \emph{inference rules}, which can each be read
as a logical inference.  That is, given evidence for the premises on
the top of the rule, we may conclude the bottom of the rule.
For instance, here are two example rules:
\[
\infer{
  \GameStep{G_1}{\ctake}{G_2}{\textsf{success}}
}{ 
  G_1 \vdash \textsf{playerNear}~\textsf{flask}
  \quad
  \quad
  \textsf{playerTake}(G_1,\textsf{flask}) \equiv G_2
}
\]
\[
\infer{
  \GameStep{G}{\ctake}{G}{\textsf{failure}}
}{
  G \vdash \textsf{not}\big( \textsf{playerNear}~\textsf{flask} \big)
}
\]
The first rule formalizes the case shown in the RHS of
Figure~\ref{fig:gameloop}.
The second rule formalizes the opposite outcome, where the flask
cannot be taken.
Notably, the first rule has two premises: To be taken by the player,
it suffices to show that in the current game state~$G_1$, the flask is
near the player (first premise), and that there exists an advanced
game state~$G_2$ that results from the player taking this flask~(second premise).
In the second rule, there is only one game state~$G$, since the flask
cannot be taken and consequently, the game state does not change.

Like the syntax of player intent and game responses, these rules are
also unambiguous.
Consequently, we view these rules as a mathematical definition with an
associated strategy for constructing formal (and informal) proofs
about the game mechanics.

For instance, we can formally state and attempt to prove that for all
player intents~$\mathit{intent}$ and game worlds, $G_1$, there exists
a corresponding game world~$G_2$ and game response~$\mathit{resp}$.
That is, the statement of the following conjecture:
\[
\begin{array}{ccc}
\forall G_1, \mathit{intent}.
&
\exists G_2, \mathit{resp}.
&\GameStep
  {G_1}{\mathit{intent}}
  {G_2}{\mathit{resp}}
\end{array}
\]
Using standard PL techniques, the proof of this conjecture gives rise
to an abstract algorithm that implements the game mechanics by
analyzing each possible case for the current game-state and player
intent.
Indeed, this is precisely the reasoning required to show that an
implementation of the game is \emph{complete} (i.e., there exists no
state and input that will lead the game into an undefined situation).

Reasoning about this completeness involves reasoning about when each
rule is applicable.  
For instance, the rule for a successful player intent of~\textsf{take}
requires two premises:
$G_1 \vdash \textsf{playerNear}~\textsf{flask}$
and
$\textsf{playerTake}(G_1,\textsf{flask}) \equiv G_2$.
The first is a \emph{logical judgment} about the game world involving
the proposition~$\textsf{playerNear}~\textsf{flask}$, which may or
may not be true, but which is computable.
The second is a \emph{semantic function} that transforms a game state
into one where the player takes a given object; in general, this
function may be undefined, e.g., if the arguments do not meet the
function's \emph{pre-conditions}.
For instance, a precondition of the
function~$\textsf{playerTake}(-,-)$ may be that the object is not
already in the possession of the player.
In this case, to show that the game mechanics do not ``get stuck'', we
must show that~$\textsf{playerNear}~\textsf{flask}$ implies that the
player does not already posses the flask.
(Otherwise, we should add another rule to handle the case that the
player intends to take the flask but already possesses it).
The design process for programming language semantics often consists
of trying to write examples and prove theorems, and failing; these
experiences inform systematic revisions to the language definition.

In the next section, we refine intents and semantics further by
introducing the notion of a \emph{context}.

\OmitThis{
\section{Precise player-game dynamics via operational semantics}
\label{sec:opsem}

\newcommand{\Player}[4]{\ensuremath{{#1}{:}\left<{#2},{#3},{#4}\right>}}
\newcommand{\PropIsTrue}[2]{\ensuremath{#1 \vdash #2~\textsf{true}}}
\newcommand{\StepsTo}[4]{\ensuremath{\left<{#1};{#2}\right> \longrightarrow \left<{#3};{#4}\right>}}

\section{Precise player-game dynamics via operational semantics}
\label{sec:opsem}




\begin{figure*}
\small

\begin{tabular}{l|l}

\begin{minipage}{0.68\textwidth}
\begin{tabular}{lccll}
unique IDs & $i,j$ & $::=$ & $\cdots$ & (\emph{abstract}, e.g., numbers, symbols, noun phrases)
\\
propositions   & $P$ & $::=$ & $\cdots$ & Game world propositions~(See~$\PropIsTrue{G}{P}$)
\\
player command & $C$ & $::=$ & $\cdot$ $~|~$ $c$ & Unit and atomic commands (0 and 1 turns, resp.)
\\
               &     & $~|~$ & $C_1~{;}~C_2$ & Command sequences (multi-turn commands)
\\
               &     & $~|~$ & $\textsf{until}~P~\textsf{do}~C$ & Command loops (e.g., for player skills)
\\
atomic command & $c$ & $::=$ & $\textsf{grasp}~i$ & Attempt to place object~$i$ into player's hand
\\
               &     & $|$   & $\textsf{drop}$ & Drop the held object, if any
\\
               &     & $|$   & $\textsf{giveTo}~i$ & Give held object to another~(named~$i$)
\\
               &     & $|$   & $\textsf{takeFrom}~i$ & Take hold of object from another~(named~$i$)
\\
               &     & $|$   & $\textsf{moveTo}~i$ & Move adjacent to object, place or agent~(named~$i$)
\\
               &     & $|$   & $\textsf{moveThrough}~i$ & Move through opening~(named~$i$)
\\[5mm]
\end{tabular}
\begin{tabular}{lccll}
game state & $G$ & $::=$ & $\cdots$ & (\emph{abstract})
\\
player state & $p$ & $::=$ & $\Player{i}{B}{h}{C}$ & Player identity~$i$, bag~$B$, hand~$h$ and commands~$C$
\\
bag state (spaces) & $B,S$ & $::=$ & $\cdot~|~s::S$ & Zero or more unit spaces
\\
hand state (unit space) & $h,s$ & $::=$ & $\square~|~i$ & An empty space, or an object~(named~$i$)
\\
\end{tabular}
\end{minipage}

&

\begin{minipage}{0.3\textwidth}
\[
\begin{array}{ll}
\fbox{$G_1 \equiv G_2$}~\textrm{World equivalence}
\\
\fbox{$B_1 \equiv B_2$}~\textrm{Bag equivalence}
\\
\fbox{$\PropIsTrue{G}{P}$} 
\\
\textrm{In world~$G$, proposition~$P$ is true}.
\\[2mm]
\fbox{$\StepsTo{G}{p}{G'}{p'}$}
\\[2mm]
\infer{  
  \StepsTo
      {G}{ 
        \Player{i}{B_1}{\square}{\textsf{grasp}~j}
      }
      {G}{ 
        \Player{i}{B_2}{j}{\cdot}        
      }
}
{
  B_1 \equiv j :: B_2
}
\\[2mm]
\infer{  
  \StepsTo
      {G}{ 
        \Player{i}{B_1}{j}{\textsf{grasp}~k}
      }
      {G}{ 
        \Player{i}{B_3}{j}{k}{\cdot}
      }
}
{
  B_1 \equiv k :: B_2
  \\
  B_3 \equiv j :: B_2
}
\\[2mm]
\infer{  
  \StepsTo
      {G}{ 
        \Player{i}{B_1}{j}{\textsf{grasp}~k}
      }
      {G}{ 
        \Player{i}{B_3}{j}{k}{\cdot}
      }
}
{
  B_1 \not\equiv k :: B_2
  \\
  B_3 \not\equiv j :: B_2
}
\end{array}
\]
\end{minipage}

\end{tabular}

\[
\begin{array}{l}
\\[2mm]
\infer{  
  \StepsTo
      {G}{ 
        \Player{i}{B}{k}{\textsf{grasp}~k}
      }
      {G}{ 
        \Player{i}{B}{k}{\textsf{errMsg}~\textrm{``you are already holding $k$''}}
      }
}
{
\quad
}
\\[2mm]
\infer{
  \StepsTo
      {G_1}{
        \Player{i}{B_1}{j}{\textsf{drop}}
      }
      {G_2}{
        \Player{i}{B_3}{\square}{\cdot}
      }
}
{
  G_2 \equiv \textsf{dropItem}(G_1, j, \textsf{pos}(i))
}
\\[2mm]
\infer{
  \StepsTo
      {G_1}{
        \Player{i}{B}{k}{\textsf{giveTo}~j}
      }
      {G_2}{
        \Player{i}{B}{\square}{\cdot}
      }
}
{
  G_1\,{\vdash}\,\textsf{isNear}(i, j)
  \\
  G_2 \equiv \textsf{giveItemTo}(G_1, k, j)
}
\end{array}
\]

\caption{Definitions for an operational semantics: Captures precise
  player-game dynamics for a primitive adventure game.}
\end{figure*}
}

  \section{Contextual interfaces as \\ type systems}
  \label{sec:typesys}

  Decisions about interface syntax can, to some extent, limit or expand the
  player's ability to form intents that will be met with failure, such as
  moving through a wall or taking an object that does not exist. But
  sometimes, whether an utterance is {\em meaningful} or not will depend on
  the runtime game state, and can be considered a distinct question from whether it is
  well-formed. For example, whether or not we can {\em take
  flask} depends on whether the flask is present, but if the flask is an
  object somewhere in the game, we must treat this command as well-formed {\em
  syntax} and relegate its failure to integrate with the runtime game
  environment to the {\em mechanics} (operational semantics).

  However, some user interfaces nonetheless restrict the set of recognized
  utterances in a way that depends on current game state. Consider a
  point-and-click interface that changes the shape of the cursor to a hand
  whenever it hovers over an interactable object, and only recognizes
  clicks when it is in this state. Alternatively, consider the hypertext
  interface, which only recognizes clicks on links made available in the
  current page.  Providing the player with {\em only the option of saying}
  those utterances that ``make sense'' in this regard corresponds to a
  strong static type system for a programming language.

  Type systems are typically formalized be defining a relation between
  expressions $e$ and contexts $\Gamma$. Contexts are sets of specific
  circumstances in which an expression is valid, or well-typed. Usually,
  these circumstances have to do with {\em variables} in the program. For
  example, the program expression $x+3$ is only well-typed if $x$ is
  a number. ``$x$ is a number'' is an example of a fact that would be
  contained in the context. Its well-typedness could be represented as
  $x{:}\mathsf{num} \vdash x+3{\ \mathsf{ok}}$.
  
  In the move-take game, we can include aspects of game state in our
  context, such as the location of the player and the adjacency mapping
  between rooms in the world. An example of a typing rule we might include
  to codify the ``only present things are takeable'' rule would be:
  \[
    \infer
    { 
      \Gamma \vdash \ctake\param{O}\ \syn{ok}
    }
    {
      \Gamma \vdash \syn{playerIn}(R)
      \quad
      \Gamma \vdash \syn{at}(O, R)
    }
  \]
  We then need to define a relation between concrete game states $G$ and
  abstract conditions on those states, $\Gamma$. We might write this
  relation $G\ :\ \Gamma$.
  After such rules are codified, we can refine the ``game completeness''
  conjecture to handle only those utterances that are well-typed:
  \[
  \begin{array}{cc}
  \forall G_1, \mathit{intent}.
  & (G_1\ :\ \Gamma)
    \land
    (\Gamma \vdash intent\ \syn{ok})
  \\
\Longrightarrow
  \exists G_2, \mathit{resp}.
  &\GameStep
    {G_1}{\mathit{intent}}
    {G_2}{\mathit{resp}}
  \end{array}
  \]
  This is nearly what we want to know about our game mechanics.
  However, we want to apply this reasoning iteratively as the game progresses, so that we
  reason next about the player intention that leads from game state~$G_2$
  to another possibly different game state~$G_3$; but what context for player intent describes state~$G_2$?
  
  For this reasoning to work, we generally need to update the original context~$\Gamma$,
  possibly changing its assumptions, and creating~$\Gamma'$.  We write $\Gamma \subseteq \Gamma'$ to mean that $\Gamma'$ \emph{succeeds} $\Gamma$ in a well-defined way. 
  Given that state~$G_1$ and~\ensuremath{\mathit{intent}} agree about the context of assumptions~$\Gamma$, we wish to show that there exists a successor context~$\Gamma'$ that agrees with the new game state~$G_2$:
  \[
  \begin{array}{lll}
&
  \forall G_1, \mathit{intent}.
  & (G_1\ :\ \Gamma)
    \land
    (\Gamma \vdash intent\ \syn{ok})
\\
\Longrightarrow &
 \exists \Gamma', G_2, \mathit{resp}.
  &
  \big( \GameStep
    {G_1}{\mathit{intent}}
    {G_2}{\mathit{resp}}
  \big)
\\
&&
\wedge 
  (G_2  :\ \Gamma') 
\wedge 
  (\Gamma \subseteq \Gamma') 
  \end{array}
  \]
  This statement closely matches the usual statement of
  \emph{progress} for programming languages with sound type systems.

  \section{Play traces as \\ straight-line programs}
  \label{sec:traces}


    If we consider the analogy of game interfaces as programming languages,
    the natural question arises, what is a {\em program} written in this
    programming language? We want to at least consider individual, atomic
    player actions to be complete programs; the preceding text provides
    such an account. But typical programs are more than one line
    long---what does it mean to sequence multiple actions in a game
    language?

    In a typical account of an imperative programming language, we
    introduce a sequencing operator $;$ where, if $c_1$ and $c_2$ are
    commands in the language, then $c_1;c_2$ is also a command.
    The operational semantics of such a command involves the
    composition of transformations on states $\sigma$:
    \[
      \infer{
        \left< \sigma; (c_1;c_2)\right> \longrightarrow
        \sigma_2
      }
      {
        \left<\sigma; c_1 \right> \longrightarrow
          \sigma_1
        \quad
        \left<\sigma_1; c_2 \right> \longrightarrow
          \sigma_2
      }
    \]
    However, interactive software makes this account more complicated by
    introducing the program response as a component. Instead of issuing
    arbitrary commands in sequence, the player may wait for a response or
    process responses in parallel with their decisions. In this respect, a
    player's ``programming'' activity more closely resembles something like
    live-coding than traditional program authoring. Execution of code
    happens alongside its authorship, interleaving the two activities. If
    we consider a round-trip through the game loop after each individual
    command issued, then what we arrive at is a notion of program that
    resembles a {\em play trace}: a log of player actions and game
    responses during a play session, e.g.
    \begin{quote}
      PLAYER: go north\\
      GAME: failure\\
      PLAYER: take flask\\
      GAME: success\\
      PLAYER: go south\\
      GAME: success
    \end{quote}
    Depending on the richness of our internal mechanics model, this play
    trace may contain useful information about changes in internal state
    related to the preconditions and effects of player actions. But the
    main important thing to note is that, despite the informal syntax used
    to present them here, these traces do not consist of strings of text
    entered directly by the player or added as log information by the game
    programmer---they are structured terms with abstract syntax that may be
    treated to the same formal techniques of interpretation and analysis as
    any program. And this syntax is at a high level of game-relevant
    interactions, not at the level of hardware inputs and engine code.

    Researchers in academia and the games industry alike have recently
    been increasingly interested in play trace data for the sake of
    analytics, such as understanding how their players are interacting with
    different components of the game and responding to this information with
    updates that support player interest~\cite{el2013game}. 
    For the most part, this trace data is collected through telemetry or
    other indirect means, like game variable monitoring, after which it
    must be analyzed for meaning~\cite{Canossa2013}. More recently,
    systems of structured trace terms that may be
    analyzed with logical queries have been
    proposed~\cite{osborn2015playspecs}, identifying as a benefit an
    ability to support automated testing at the level of design intents.
    Our PL analogy supports this line of inquiry and warrants
    further comparison and collaboration.

%
%
%


\section{Player skills as \\ general programs}
\label{sec:skills}

While considering ``straight-line'' traces may have some utility in player
analytics, a more exciting prospect for formalizing game interactions as
program constructs is the possibility of
encoding {\em parameterized} sequences of actions that may carry out
complex tasks. After all, games with rich player action languages afford
modes of exploratory and creative play: consider item crafting in
Minecraft~\cite{minecraft}, puzzle solving in Zork~\cite{blank1980zork}, or
creating sustainable autonomous systems like a supply chain in
Factorio~\cite{factorio}, a farm in Stardew Valley~\cite{stardew}, or a
transit system in Mini Metro~\cite{minimetro}.  Each of these activities
asks the player to understand a complex system and construct multi-step
sequences of actions to accomplish specific tasks. From the player's
perspective, these plans are constructed from higher-level activities,
such as {\em growing a crop} or {\em constructing a new tool}, which
themselves are constructed from the lower level game intent language.

A language, as we have formalized it, gives us the atomic pieces from which
we can construct these sequences, like Lego bricks can be used to construct
reconfigurable components of a house or spaceship. {\em Compositionality}
in language design is the principle that we may understand the meaning and
behavior of compound structures (e.g. sequences) in terms of the meaning
and behavior of each of its pieces (e.g. actions), together with the
meaning of how they are combined (e.g. carried out one after the other, or
in parallel). In this section, we describe how we might make sense of {\em
player skills} in terms of programs written in a more complex version of
the player language.

\begin{figure}
  \includegraphics[width=0.45\textwidth]{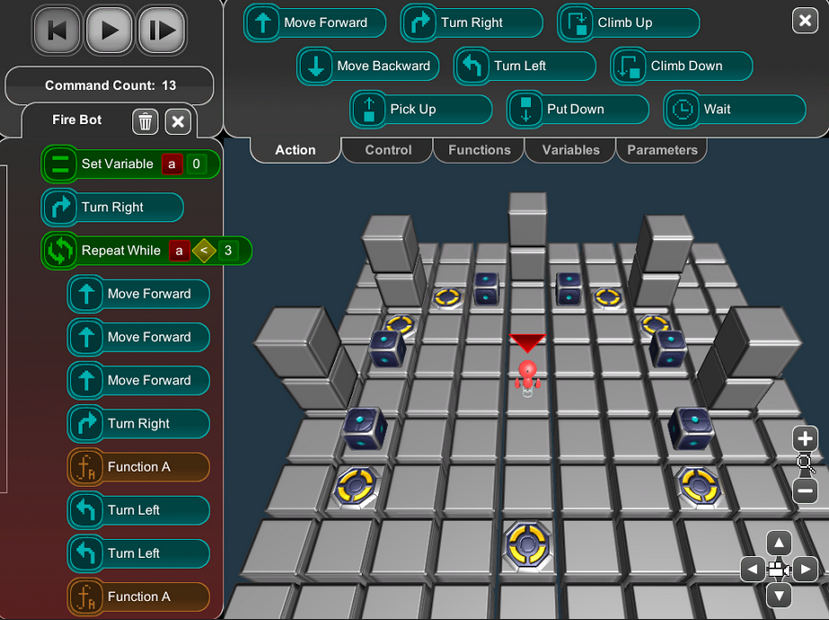}
  \caption{A screenshot from BOTS, an educational game in which players
  write programs to direct a player avatar.}
  \label{fig:bots}
\end{figure}

Such programs might be integrated into a game's mechanics so that a player
explicitly writes such programs, as in the BOTS game, an interactive
programming tutor that asks players to write small imperative programs that
direct an avatar within a virtual world~\cite{hicks2012creation} (see
Figure~\ref{fig:bots}), or Cube
Composer\footnote{\url{http://david-peter.de/cube-composer/}}, in which
players write functional programs to solve puzzles. However, for now, we
primarily intend this account of player skills as a conceptual tool.

\subsection{Example: Stardew Valley}

\begin{figure}
  \includegraphics[width=0.45\textwidth]{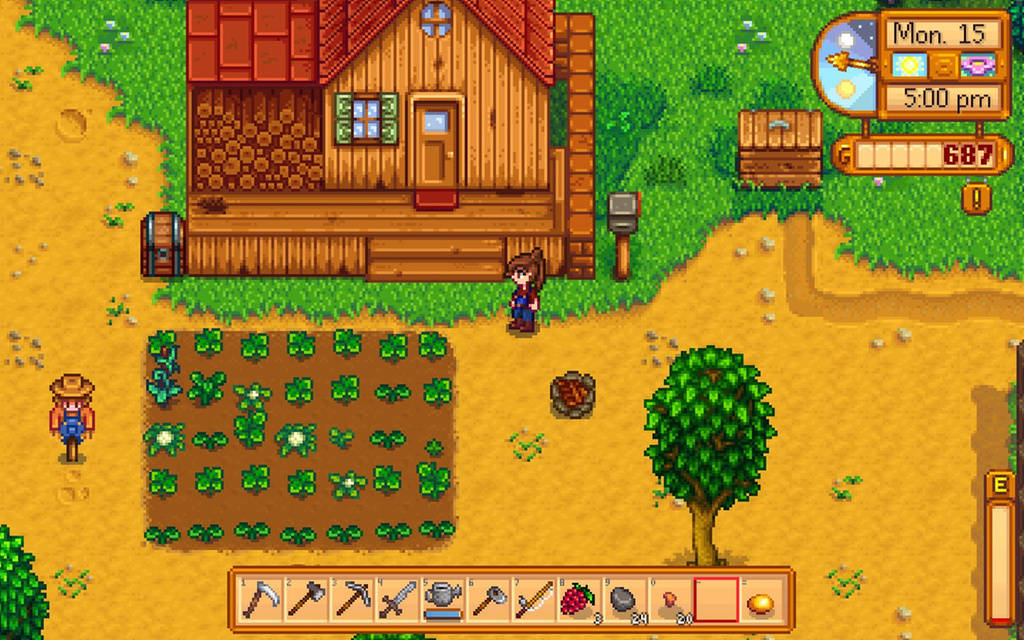}
  \caption{A screenshot from Stardew Valley showing the player's farm,
  inventory, and avatar.}
  \label{fig:stardew}
\end{figure}

Our initial $\{\cmove, \ctake\}$ example is too simple to craft really
compelling examples of complex programs, so here we examine
Stardew Valley and its game language for the sake of considering player
skills. In Stardew Valley, the player has an inventory that permits varied
interactions with the world, beginning with a number of tools for farming (axe,
hoe, scythe, pickaxe) which do different things in contact with the
resources in the surrounding environment; most include extracting some
resource (wood, stone, fiber, and so on), which themselves enter the
player's inventory and can be used in further interaction with the game
world. The player's avatar is shown on-screen, moved by WASD.  There are
also context-sensitive interactions between the player and non-player
characters (NPCs), interfaces through which new items may be purchased
(shops), and mini-games including fishing (fish may also be sold at high
value). See Figure~\ref{fig:stardew} for a typical player view of the game.

While a full account of the language that this game affords the player is
beyond the scope of this paper, we include a representative sample of the
actions and affordances found in this game that may be used to construct
player skills.
These include: directional avatar movement within and between
world ``rooms,'' point-and-click actions for selecting items in
one's inventory, and interacting with in-room entities. 

The player avatar must be near an entity for the player to interact with
it. They can then either $\syn{apply}$ the currently-selected inventory
item to the in-world entity with a left-button click, or right-button
click, which does something based on the entity type, e.g. doors and chests
open, characters speak, and collectible items transfer to the player's
inventory. We refer to this last action as $\syn{inquire}$. We also note
that, for the sake of our example, movement towards an entity and movement
offscreen (towards another room) are the only meaningful and distinct types
of movement, which we refer to as $\syn{move\_near}$ and
$\syn{move\_offscreen}$. These actions yield the following syntax:
%
\begin{eqnarray*}
intent &::=& \syn{select} \param{item}\\
       &\mid& \syn{apply} \param{entity}\\
       &\mid& \syn{inquire} \param{entity}\\
       &\mid& \syn{move\_near} \param{entity}\\
       &\mid& \syn{move\_offscreen} \param{direction}
\end{eqnarray*}
We leave the definition of items and entities abstract, but we could
imagine it to simply list all possible items and entities in the world as
terminal symbols.  
From these atomic inputs, we can start to construct higher-level actions
performed in the game most frequently---tilling land, planting seeds,
conversing with NPCs, and so on. These blocks of code may be assigned names
like functions to be called in many contexts:
%
%
%
\begin{verbatim}
action till  = select hoe; move_near hard_ground; 
               apply hard_ground
action plant = select seeds; move_near tilled_ground; 
               apply tilled_ground
action mine  = select pickaxe; move_near rock; apply rock
action talk  = move_near npc; inquire npc
action enter_shop = move_near shop; inquire door(shop)
\end{verbatim}
In turn, these larger skill molecules may be combined to accomplish
specific tasks or complete quests.

\subsection{Branching programs as skills and strategies}

Note that we have naively sequenced actions without consideration for the
game's response. This approach to describing player skills does not take
into account the possbility of a failed attempt, such as attempting to mine
when there are no rocks on the current screen. One could simply assign a
semantics to these sequences of action that threads failure through the
program---if we fail on any action, the whole compound action fails.

However, we can go further with describing robust player skills and
strategies if we consider the possibility of {\em handling} failure, a
common feature of day-to-day programming and indeed of gameplay. Recall our
simplified game response language consisting of two possibilities,
$\syn{success}$ and $\syn{failure}$. We can introduce a $\syn{case}$
construct into our language to handle each of these possibilities as a
distinct branch of the program:
\begin{verbatim}
action mine = 
  response = select pickaxe;
  case(response):
    success => {
      response' = move_near rock;
      case(response'):
          success => apply rock;
        | failure => fail;
    }
  | failure => fail;
\end{verbatim}
However, to avoid handling failure at every possible action, a better
approach is to explicitly indicate as parameters the world resources that
each action needs in order to complete successfully.
The overall action definition for mining, or example, would
require as a precondition to the action that a pickaxe is
available in the player inventory and a rock is in the same room. The
actions for selecting the pickaxe and moving near the rock would depend on
these resources, and the game response language could include the resources
it guarantees as outputs. Then we can write the program using simpler
notation that refers to resource dependencies of the appropriate type
(using notation \verb|resource : type|):
\begin{verbatim}
action mine(p:pickaxe, r:rock) = 
  select p; move_near r; apply r
: mineral
\end{verbatim}
%

This notation together with a branching \verb|case| construct
scales to include nondeterminism in the game world, such as the fishing
minigame in Stardew Valley: the game always eventually tells the player
that something is tugging at her line, but some portion of the time it is a
fish while the rest of the time it is trash.  These constructs can also
account for incomplete player mental models, such as knowing that one must
water her seeds repeatedly day after day in order to grow a crop, but not
knowing how many times.

Below, we present a notation that accounts for these aspects of player
skills: a \verb|do ... recv ...| notation indicates a command and then
binds the response to a {\em pattern}, or structured set of variables,
which can then be case-analyzed.
Our first example is watering a crop until it may be harvested:
\begin{figure}[t]
  \includegraphics[width=0.4\textwidth]{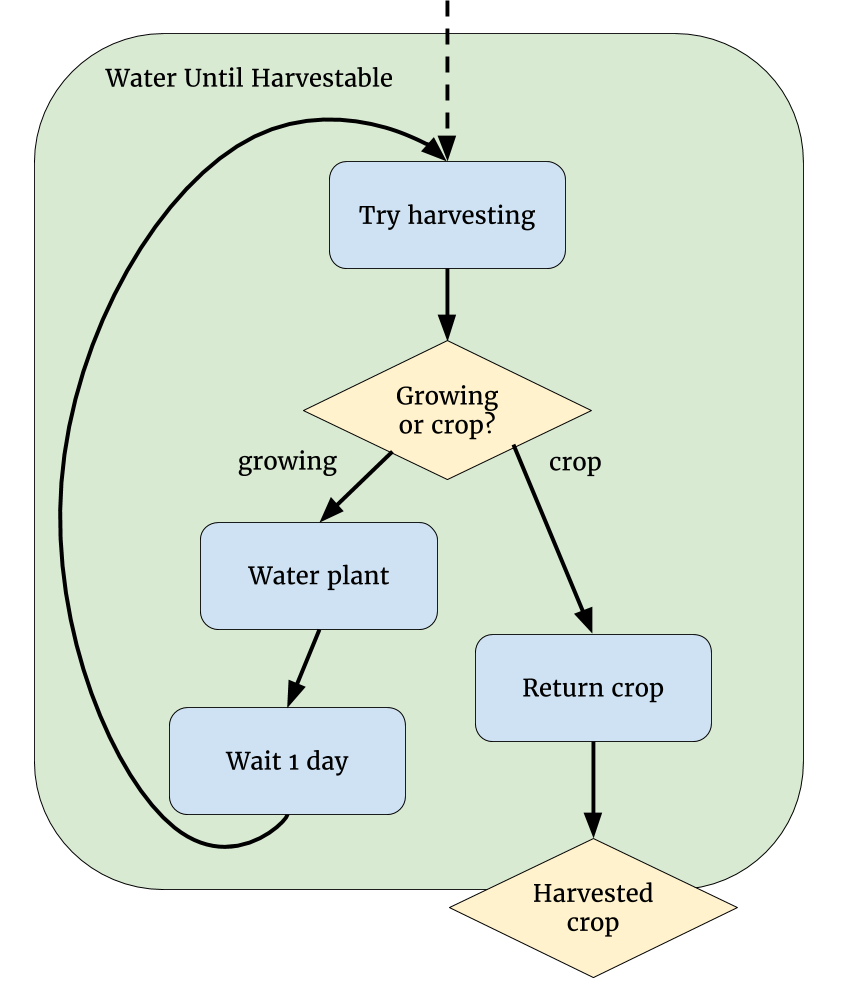}
  \caption{Watering a crop until it may be harvested.}
  \label{fig:harvest}
\end{figure}
\begin{verbatim}
action water_until_harvestable[t](p: planted(t))
: crop(t) =
  do try_harvest(p)
    recv <result: crop(t) + growing(t)>.
      case result of
        c:crop(t) => c
      | g:growing(t) => 
          water(g); 
          wait(day); 
          try_harvest(g)
\end{verbatim}
See Figure~\ref{fig:harvest} for a control flow diagram of this code.

The next example shows a parallel construct \verb/||/, which can be used to
compose actions with distinct dependencies, as well as
how an action definition may use other action definitions by threading
resource dependencies through as arguments:
\begin{figure}[t]
  \includegraphics[width=0.4\textwidth]{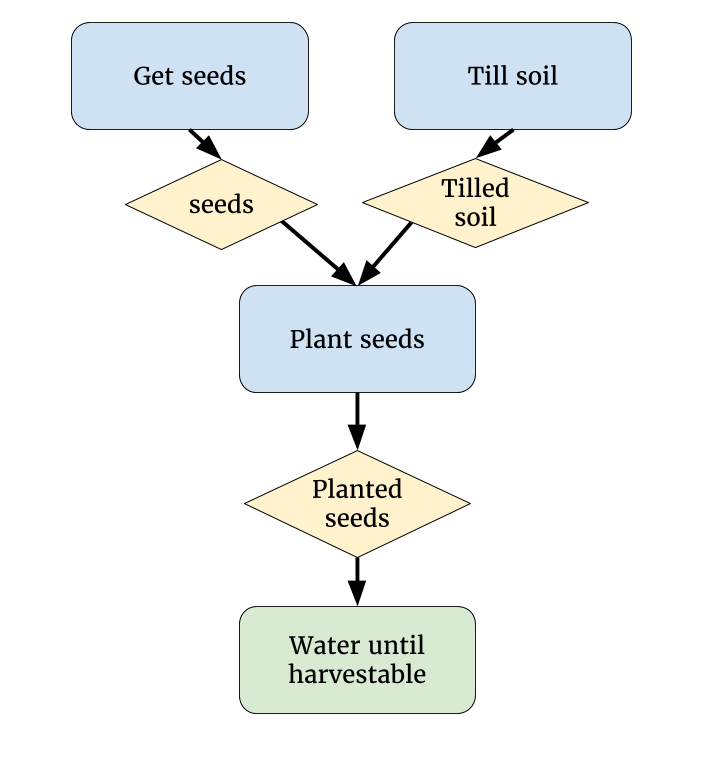}
  \caption{Growing a crop.}
  \label{fig:grow}
\end{figure}
\begin{verbatim}
fun grow_crop[t : croptype](s:soil, w:watering_can)
: crop(t) =
  do
    get_seeds(t) || till_soil(s)
  recv <s: seeds(t), g: tilled_soil>.
    do
      plant(s, g)
    recv <p: planted(t)>.water_until_harvestable(p)
\end{verbatim}
See Figure~\ref{fig:grow} for a control flow diagram of this code.

\section{Discussion}
\label{sec:discussion}

Having established a vocabulary of syntax and semantics for player
languages, we can now revisit the potential benefits of this account
mentioned in the introduction and discuss them in more detail.

\subsection{Composing play traces, player skills}

One of the major things that a programming language account provides is
{\em compositionality}: a system for making sense of meaning of a complex
artifact in terms of the meaning of its pieces. This comes up in two places
for looking at games:

{\bf Structured play traces.} With a formalized game language, the sequence
of steps along the transition system described by the operational semantics
forms a mathematical artifact that is subject to deeper analysis than what
can be gained simply from screen or input device recordings. For example,
we can carry out causal analysis, asking ``why'' queries of trace data,
e.g. ``Why was the player able to unlock the door before defeating the boss?'',
as well as filtering traces for desired properties: ``Show me a play trace
where the player used something other than the torch to light the room.''
The recent PlaySpecs project~\cite{osborn2015playspecs} suggests interest
in formulating traces this way to support this kind of query.

{\bf Player skills as programs.} While a play trace may be interpreted as a
straight-line program, even more interesting is the idea of latent
structure in player actions, such as composing multiple low-level game
actions into a higher-level skill, following the cognitive idea of
``chunking.'' We map this idea onto that of {\em functions} in the
programming language that take arguments, generalizing over state space
possibilities (e.g.: the red key opened the red door, so for all colors
$C$, a $C$ key will open a $C$ door). Further reasoning forms like case
analysis to handle unpredictable game behavior and repeating an action
until a condition holds are also naturally expressed as programming
language constructs.

\subsection{Abstraction boundaries \\ between input and mechanics}


Formalizing game interfaces gives us the tools to explore alternative
interfaces to the same underlying mechanics, without needing to port
game logic between different graphical interface frameworks. For example,
the interactive fiction community has been exploring alternatives to the
traditional dichotomy of ``parser vs. hypertext'' for presenting
text-based games and interactive story-worlds. An abstraction boundary
between the underlying mechanics, map, and narrative of the world, and the
view and input mechanisms used to interact with it, could open the doors to
research on user interfaces that support players' mental models of a world
conveyed in text.

\subsection{Enabling co-creative play}

Finally, a PL formulation of player actions along with appropriate
composition operators (parallel and sequential composition, branching, and
passing resource dependencies) provides a ``scripting language for free''
to the game environment. 
Such a language can be used to test the game, provided as a game mechanic
as in BOTS, or provided as an optional augmentation to the game's mechanics
for the sake of modding or adding new content to the game world. Especially
in networked game environments, like multi-user domains, massively
multiplayer online games, and social spaces like Second Life, the ability
for the player to program not just her avatar but parts of the game world
itself introduces new opportunities for creative and collaborative play.
Our framework suggests a new approach to designing these affordances for
players in a way that is naturally derived from the game's existing
mechanics and interface.  In the spirit of {\em celebrating the player},
this year's conference theme, we wish to enable the player as a
co-designer of her own game experience.

\subsection{Future Work}
  In future work, we intend to build software for realizing game language
  designs and experimenting with protocol-based game AI developed as
  programs in these languages. In another direction, we aim to innovate in
  PL design outside of games, such as read-eval-print loops (REPLs) for
  live programming that includes rapid feedback loops motivated by
  gameplay, as well as distributed and concurrent systems that may benefit
  from the protocol-based approach proposed here.

\section{Conclusion}
\label{sec:conclusion}

We propose \emph{player intent languages}, a systematic framework for
applying programming language design principles to the design of
player-game interfaces.
We define this framework through simple examples of syntax (aka player
intents), type systems (aka contextual interfaces) and operational
semantics (aka game mechanics).
We show how applying this framework to a player-game interface
naturally gives rise to formal notions of play traces (as
straight-line programs) and player skills (as general programs with
branching and recursion).
By defining a player intent language, game design concepts become
formal objects of study, allowing existing PL methodology to inform and
guide the design process.

%
%
  
%
%
%


\bibliographystyle{ACM-Reference-Format}
\bibliography{main,adapton} 

\end{document}